\documentclass[referee]{raa}            
\usepackage{graphicx}
\usepackage{subcaption}  
\usepackage{float}       
\usepackage{placeins}    
\usepackage{siunitx}
\usepackage{graphicx,times}             
\usepackage{natbib}
\usepackage{amssymb,amsmath}
\bibpunct{(}{)}{;}{a}{}{,}
\usepackage{comment}
\usepackage{placeins}

\begin{document}

\title{{Design and Validation of the Digital Receiver System for the next-generation radio interferometer}}

   \volnopage{Vol.0 (20xx) No.0, 000--000}      
   \setcounter{page}{1}          

   \author{Donghao Qu \footnote{dhqu@shao.ac.cn}
      \inst{1,2}
   \and Jiajun Zhang \footnote{jjzhang@shao.ac.cn}
      \inst{1,3}
   \and Yajun Wu
      \inst{1,3}
   \and Zhang Zhao
        \inst{1}
   \and Zixuan Liu
        \inst{1,2}
   \and Yanbin Yang
        \inst{1}
   }

\institute{
\textsuperscript{1}Shanghai Astronomical Observatory, Chinese Academy of Sciences, 80 Nandan Road, Shanghai 200030, China; \\
\textsuperscript{2}School of Astronomy and Space Science, University of Chinese Academy of Sciences, No. 19A Yuquan Road, Beĳing 100049, China;\\
\textsuperscript{3}State Key Laboratory of Radio Astronomy and Technology, 20A Datun Road, Chaoyang District, Beijing, 100101, China
}
\vs\no
   {\small Received 20xx month day; accepted 20xx month day}

\abstract{{This paper presents the design and validation of a digital receiver system developed for the next-generation radio interferometer projects. The receiver supports 8 analog inputs with 12-bit, 4GHz sampling and performs real-time signal processing using FPGA-based channelization. Field experiments were conducted to observe the Sun, a satellite beacon, and Cassiopeia A. Interference fringes were analyzed and modeled. Time delay compensation was implemented in two ways: theoretical calculation and Gaussian Process Regression (GPR) fitting. Results show sub-nanosecond consistency between the two methods. The field experiments demonstrate the receiver’s suitability for future radio telescopes such as the BINGO-ABDUS project.}}

   \authorrunning{Donghao Qu, Jiajun Zhang \& Yajun Wu }            
   \titlerunning{DH1 Digital Receiver System}  

   \maketitle
%
%
\section{Introduction}           
Radio interferometry has become an essential technique in modern radio astronomy, allowing arrays of spatially separated antennas to synthesize a virtual telescope with high angular resolution \citep{interference}. By measuring the time delays and correlated signals between antenna pairs, interferometers can reconstruct fine-scale astronomical structures and conduct high-precision sky surveys. Prominent examples of radio interferometers include the
LOFAR(Low Frequency Array)\citep{2013lofar}, operating at low frequencies to study cosmic dawn and ionospheric effects; the Murchison Widefield Array (MWA)\citep{mwa}, optimized for widefield observations of the southern sky; CHIME(Canadian Hydrogen Intensity Mapping Experiment)\citep{chime}, designed for 21-cm intensity mapping to detect baryon acoustic oscillations; and the upcoming SKA\citep{skaphase}, which aims to be the most sensitive radio interferometer ever built. In China, the Tianlai project is advancing cosmological studies\citep{2021MNRAStianlai}. The BINGO-ABDUS project located in Brazil, is advancing intensity mapping and cosmology studies using hybrid or phased array technologies\citep{bingo1}.

The performance of a radio interferometric array is closely linked to the backend digital processing system. In recent years, reconfigurable FPGA-based digital receiver platforms such as ROACH2 (Reconfigurable Open Architecture Computing Hardware) \citep{roach2tianlai} and SKARAB (SKA Reconfigurable Application Board) \citep{skarab} have been widely adopted in large-scale radio telescope projects. The digital backend system of QTT is implemented using a hybrid architecture that combines Field-Programmable Gate Arrays (FPGAs) with CPUs and GPUs \citep{QTT}. These systems support high-speed analog-to-digital conversion (ADC), real-time channelization, and flexible data transmission, making them suitable for wideband and multi-element interferometry.

A key challenge in interferometric systems is the compensation of geometric and instrumental time delays between antennas. These delays arise due to differences in signal propagation paths and electronic components. If uncorrected, they cause phase decoherence, distorted fringes, and loss of sensitivity. Therefore, accurate delay estimation and compensation—both integer and sub-sample level—is critical to achieve optimal fringe contrast and dynamic range. Among the advanced techniques, Gaussian Process Regression (GPR) has shown promise in robustly fitting phase-frequency relationships for delay estimation in noisy or RFI-contaminated environments \citep{GP,RGPR}.

In Sec.~\ref{sect:bg}, we present the design and specification of a high-speed digital receiver system tailored for radio interferometric applications, in particular for the BINGO/ABDUS phased array experiment. In Sec.~\ref{sec:obs}, we present the observations we have conducted for validating the digital receiver. In Sec.\ref{sec:delay_comp}, we demonstrate delay compensation using both geometric modeling and signal-based estimation (via cross-correlation and GPR), and validate the performance through field tests involving observations of celestial and artificial sources. We conclude and discuss in Sec.~\ref{sec:conclusion}. The system architecture and compensation pipeline developed here provide a scalable solution for future interferometric arrays.

\label{sect:intro}


\section{Background}
\label{sect:bg}
\subsection{BINGO-ABDUS}

To expand BINGO's scientific capabilities, the BINGO-ABDUS has been proposed. ABDUS will enhance the original BINGO design by integrating phased array detectors and outrigger stations\citep{abdus}, significantly increasing the number of beams from 28 up to 200 through multibeam electronic synthesis. This upgrade will enable broader sky coverage, reaching up to 50\% of the sky with resolutions between 27' and 40', and extend redshift coverage up to $z \approx 2.1$. The use of phased arrays will not only improve survey speed and sensitivity but also enhance RFI(Radio Frequency Interference) mitigation, localization of transient sources, and overall data quality.

The ABDUS project envisions the deployment of Vivaldi Aperture Array Outrigger Stations (VAAOS)\citep{vivaldi}, located approximately 20 km from the main BINGO dish. These stations, composed of dense, connected arrays of tiles, are optimized for wideband operation (0.5--1.5 GHz) and high angular resolution through interferometric techniques. 

In support of the BINGO-ABDUS project's technical requirements, we conducted receiver tests focusing on signal correlation, time delay compensation, and fringe analysis. These tests verify the system's performance under simulated observation conditions and ensure readiness for the phased array and outrigger implementations. Our experimental work provides critical validation for the upcoming BINGO-ABDUS deployments, especially in the areas of beamforming accuracy, time delay calibration, and interference fringe detection essential for high-precision cosmological and transient radio observations \citep{transient}.

\subsection{BINGO interferometry system}
To enhance the capabilities of the BINGO telescope for transient science, particularly the detection and localization of Fast Radio Bursts (FRBs), the BINGO Interferometry System (BIS) has been proposed. BIS augments the single-dish BINGO setup with a set of auxiliary outrigger antennas—smaller, single-horn radio telescopes optionally equipped with mirrors of varying sizes (4 m, 5 m, or 6 m). These outriggers, placed at distances of 10–40 km from the main BINGO dish, enable interferometric observations by forming baselines between pairs of telescopes. By performing cross-correlations, BIS achieves arcsecond-scale localization of FRB events \citep{frb}.

The BIS concept was quantitatively evaluated through the development of a simulation framework named \texttt{FRBlip}, which generates cosmological mock FRB catalogs and models detection and localization capabilities. Simulation results indicate that a BIS configuration with 9 outriggers (each with a 6 m mirror) can localize over 20 FRBs per year, primarily in the redshift range $z \lesssim 1$. Wider beam outriggers increase the chances of multi-baseline detections, enhancing localization efficiency, while larger mirrors provide better sensitivity but narrower sky coverage. The BIS, therefore, represents a significant extension of BINGO’s scientific reach, enabling joint studies of cosmology and radio transients using interferometric techniques \citep{BIS}.

\subsection{digital receiver}
\begin{figure}[H]
    \centering
    \includegraphics[width=0.7\linewidth]{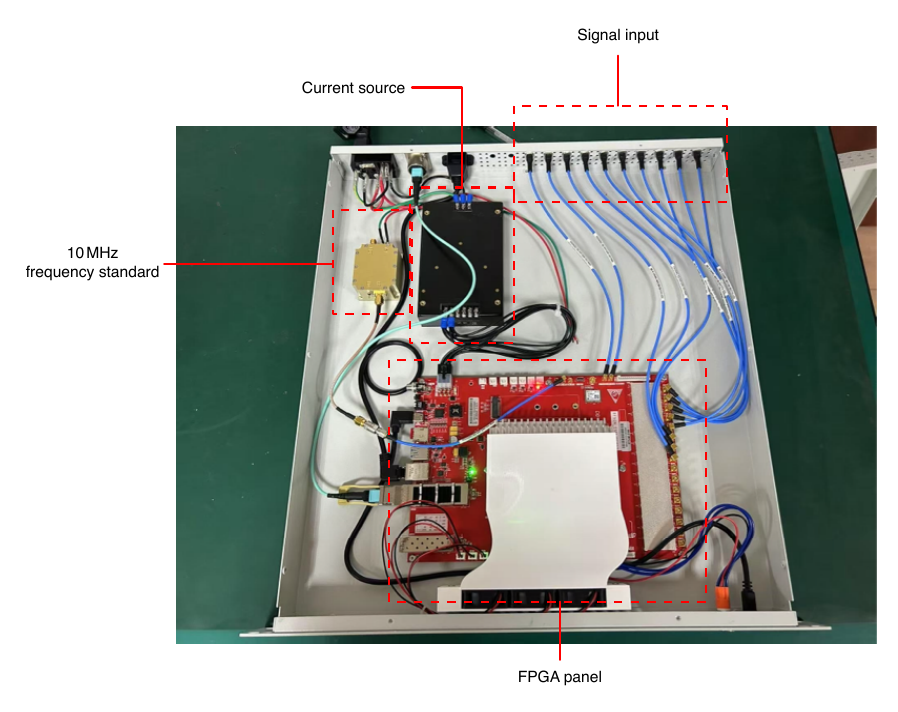}
    \caption{Block diagram of the digital receiver unit DH1. The system consists of multiple key components, including: a 10\,MHz frequency standard (left), which provides a stable reference signal for clock synchronization; a current source (top center) that powers analog front-end modules; a signal input section (top right) where multiple analog signals are received via coaxial connectors; and the main FPGA processing board (bottom), which performs digitization, channelization, and real-time signal processing. The interconnections between modules are implemented using shielded cables and coaxial lines, and the layout is optimized to minimize noise and crosstalk for sensitive astronomical data acquisition.}
    \label{fig:board}
\end{figure}

To meet the requirements of next-generation radio telescopes like BINGO-ABDUS, the digital receiver needs to have low latency, wide band, high time resolution, and a large number of channels. We have used the RFSoC system, based on Xilinx FPGA development board, to make up the digital receiver. The specifications are listed in Tab.~\ref{tab:receiver_specs}. The photograph of the digital receiver is shown in Fig.~\ref{fig:board}. We give it the name "DH1".

About the digital receiver, the 8-channel ADCs operate simultaneously at a sampling rate of 4096 MHz with a sampling bandwidth of 2048 MHz. The digitized signals are divided into $16\times128$ MHz sub-bands using a polyphase filter bank (PFB). A subset of these sub-bands is selected based on observational requirements, and the resulting data are formatted into frames using the VLBI Data Interchange Format (VDIF). These frames are then transmitted via a 100 Gbps network interface.
\begin{table}[h]
    \centering
    \caption{Specifications of the Digital Receiver Unit}
    \begin{tabular}{p{4cm} | p{8cm}}
        \hline
        \textbf{Module} & \textbf{Function and Specifications} \\
        \hline
        Sampling and Processing & 
        \begin{itemize}
            \item 8 analog input channels
            \item 12-bit resolution
            \item Maximum sampling rate: 4\,GHz
            \item 4096-point FFT in FPGA
            \item 100Gbps QSFP28 communication connector
        \end{itemize} \\
        \hline
        ASC (Analog Signal Conditioning) & 
        \begin{itemize}
            \item 8 analog inputs and 8 analog outputs
            \item Programmable gain control
            \item Anti-aliasing filters
        \end{itemize} \\
        \hline
        Timing Module & 
        \begin{itemize}
            \item 10\,MHz reference frequency distribution
        \end{itemize} \\
        \hline
        Communication and Control &
        \begin{itemize}
            \item 1\,Gb Ethernet Control interface
            \item Control software for communication with the Monitor and Control (M\&C) system
        \end{itemize} \\
        \hline
    \end{tabular}
    \label{tab:receiver_specs}
\end{table}

\section{Observation and Data Processing}
\label{sec:obs}
\subsection{Xi'an antenna array}
In this experiment, we perform observations in Xi'an. The 13-meter and 16-meter antennas were used for L-band signal correlation and interference fringe testing. The dual-polarized signal from the 13-meter antenna and the single-polarized signal from the 16-meter antenna are simultaneously fed into the data acquisition terminal. 

The data acquisition terminal collects signals from 8 analog input channels, with each channel having a bandwidth of 128 MHz, an 8-bit resolution, and a recording data rate of about 16 Gbps. In this experiment, three channels of the input signals are effective signals from the two antennas, while the other inputs are unconnected. The target sources are a satellite at 1561 MHz in the L-band and the Sun, and the 13-meter and 16-meter antennas are used for testing.

\subsubsection{Satellite Real-time Monitoring}
\begin{figure}[H]
    \centering
    \includegraphics[width=0.8\linewidth]{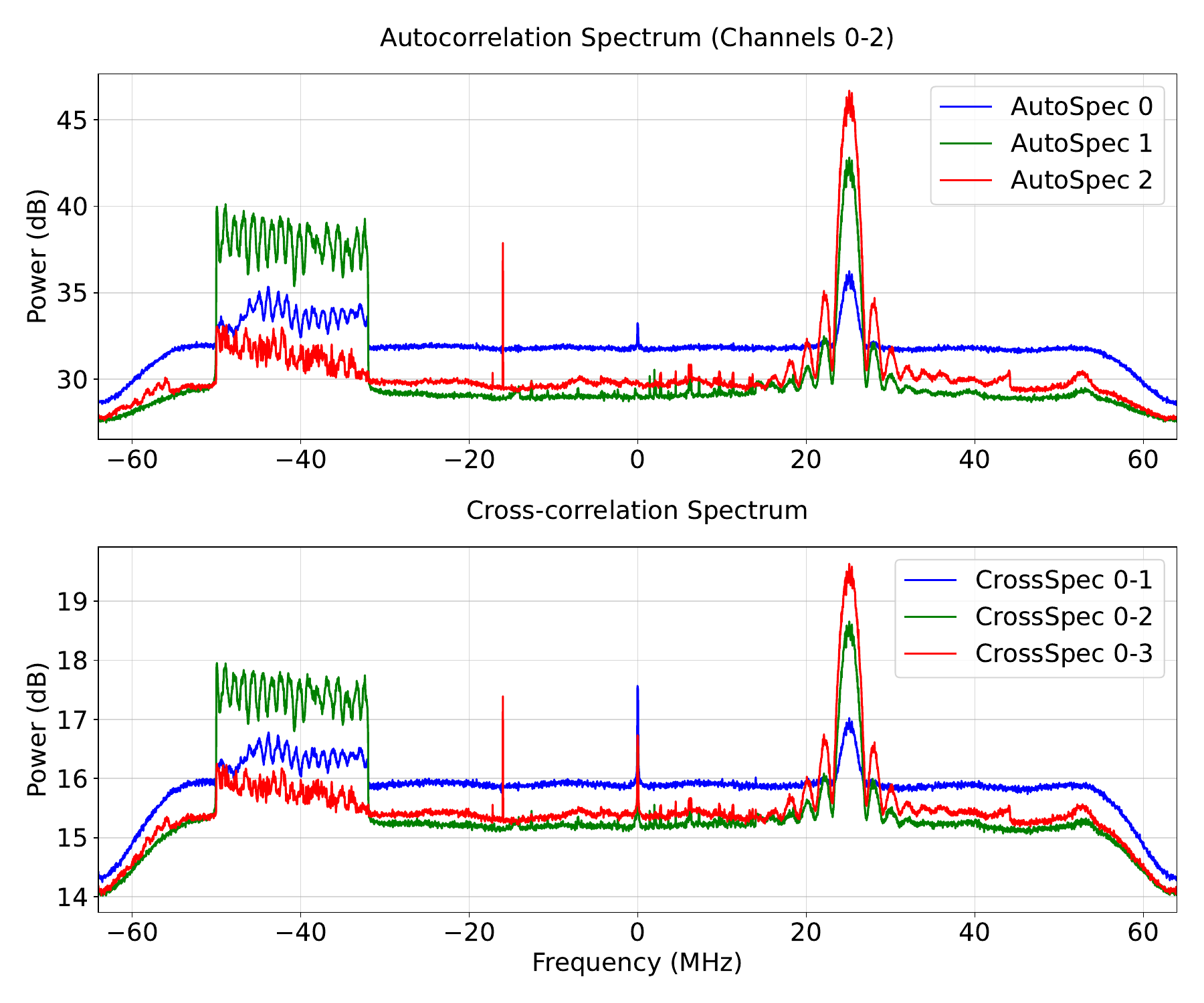}
    \caption{The upper panel represents the  Self-correlation spectra of the received signals: RCP from the 16-m antenna (red) and linear polarizations XX (blue), YY (green) from 13-meter antenna. The amplitude variations reflect frequency-dependent system response. The lower represents cross-correlation amplitudes between antenna pairs. Phase slopes (not shown) encode geometric delays relative to 1536~MHz RF center frequency. The baseband range of $\pm$64~MHz is shown.}
    \label{fig:self-correlation}
\end{figure}

During the observation, data from the acquisition terminal is captured in real-time for online processing. The antennas are pointing at a navigation satellite. The analysis includes both self-correlation and cross-correlation measurements:

Self-correlation (autocorrelation) reveals the power spectrum of individual signals, computed as:
\begin{equation}
R_{xx}(\tau) = \int x(t)x^*(t-\tau) dt
\end{equation}
where x(t) represents the signal voltage. The upper panel in Fig.~\ref{fig:self-correlation} shows the self-correlation spectra of the three signals, where the red line represents the right-circularly polarized (RCP) signal from the 16-meter antenna, while the blue and green lines represent the dual-linearly polarized (XX and YY) signals from the 13-meter antenna. The peak amplitudes indicate system gain variations across frequencies.

Cross-correlation measures the coherence between different signals:
\begin{equation}
R_{xy}(\tau) = \int x(t)y^*(t-\tau) dt
\end{equation}
The lower panel in Fig.~\ref{fig:self-correlation} illustrates the normalized cross-correlation amplitudes among the three signals. The horizontal axis represents the baseband frequency, with the center frequency of 0 MHz corresponding to a radio frequency of 1536 MHz. The phase slopes in these correlations contain geometric delay information crucial for imaging.

\subsubsection{Sun Real-time Spectrum and Correlation Monitoring}

\begin{figure}[H]
    \centering
    \includegraphics[width=0.8\linewidth]{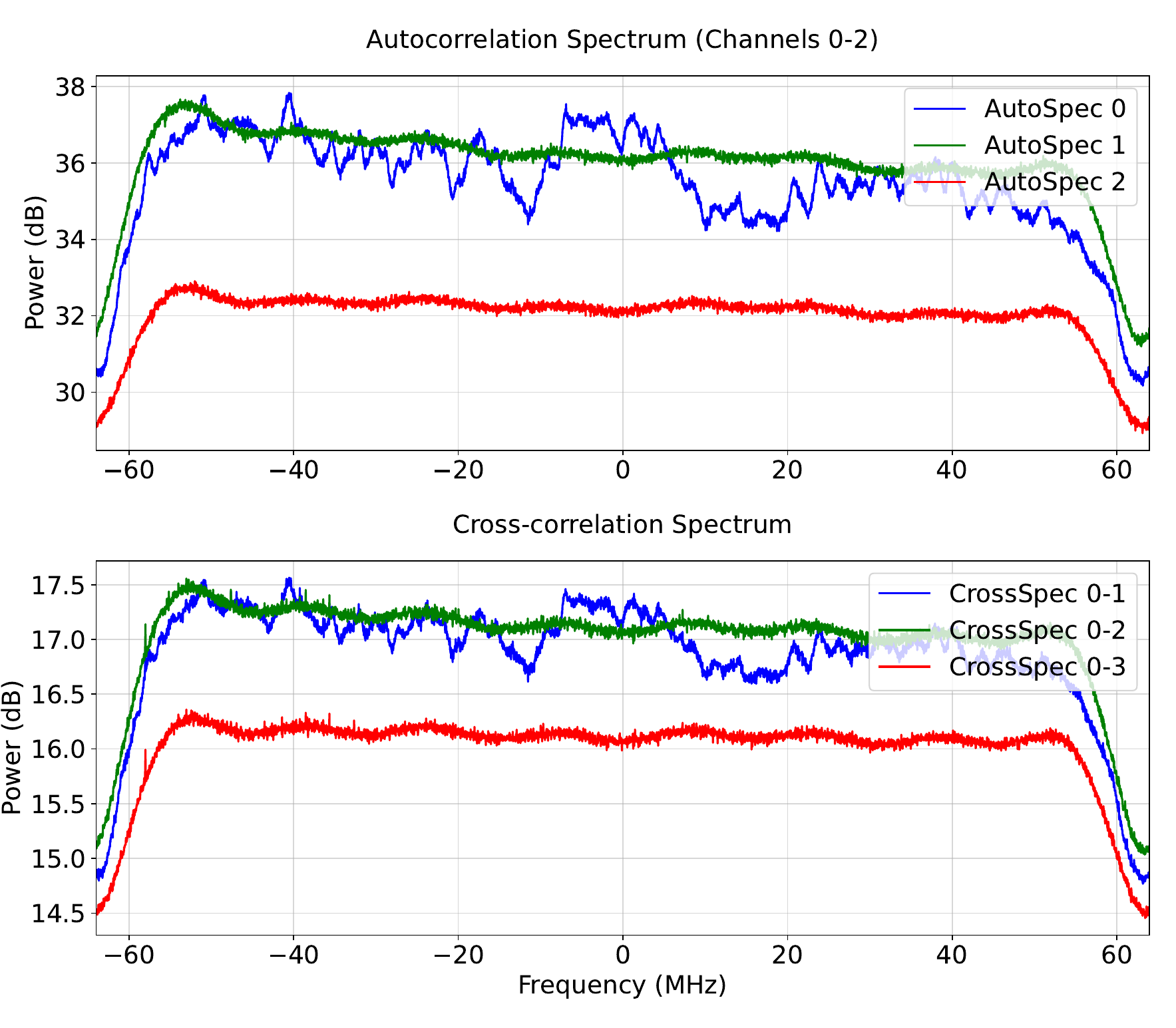}
    \caption{The upper panel is the auto-correlation spectrum of the sun, while the lower is the cross-correlation spectrum of the sun.}
    \label{fig:sun-self}
\end{figure}

During the observation, the antennas are tracing the sun. Real-time monitoring software was used to inspect the spectrum and correlation of the three signals. The upper panel in Fig.~\ref{fig:sun-self} shows the spectrum, where the red line represents the spectrum of the 16-meter antenna, the green line represents one polarization of the 13-meter antenna, and the blue line represents the other polarization of the 13-meter antenna. 

\subsection{Tianlai array}

\begin{figure}
    \centering
    \includegraphics[width=1.0\linewidth]{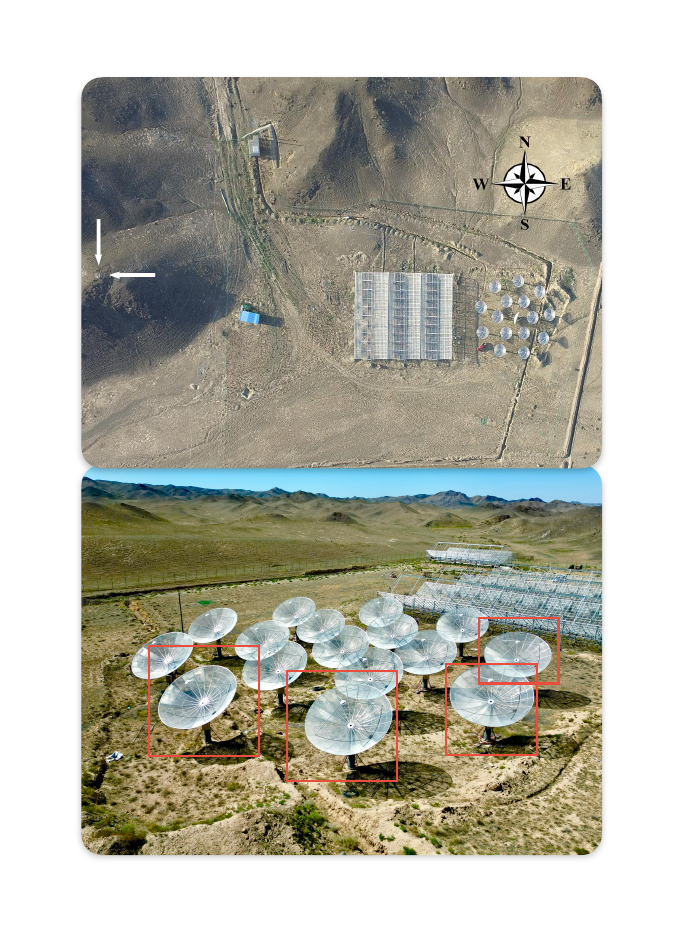}
    \caption{The upper panel is the top view of the Tianlai Dish Array Pathfinder and Cylinder Array Pathfinder taken with a DJI M600 Pro drone at a height of 280 m above the ground. The position of the calibration CNS is indicated by the white arrows on the left. The relative distance vector from the feed in dish 16 at the center of the array (when pointed toward the zenith) to the CNS is [-184.656, 13.915, 12.588] meters, with x,y,z to the east, north, and zenith. The CNS is in the far-field of all dishes in the dish array.(credit: \protect\citep{2021MNRAStianlai}).The lower panel is an aerial photograph of the Tianlai dish antenna array photoed by DJI mini 3 pro. The four antennas used for digital receiver testing are marked with red squares.}
    \label{fig:tianlai array}
\end{figure}

The experiment used four Tianlai dish antennas\citep{2021MNRAStianlai} for L-band signal correlation and interference fringe testing. Each of the four parabolic antennas has two polarized signal inputs. The antenna distribution of the Tianlai Array and the antennas used for testing are shown in Fig.~\ref{fig:tianlai array}.

\begin{figure}[H]
    \centering
    \includegraphics[width=1\linewidth]{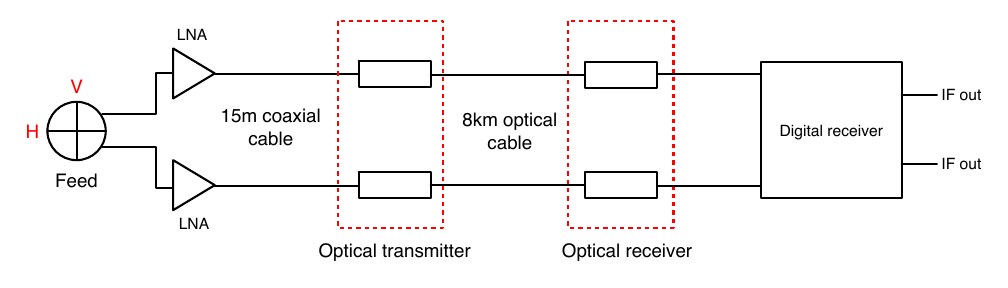}
    \caption{schematic of the RF analog system. The radio signals from the feed are separated into horizontal (H) and vertical (V) polarizations. Each polarization signal is amplified by a low-noise amplifier (LNA), and then transmitted through a 15-meter coaxial cable to an optical transmitter. The optical transmitter converts the analog RF signal into an optical signal, which is transmitted over an 8-kilometer optical fiber link to an optical receiver. The optical receiver converts the signal back to the RF domain and delivers it to the digital receiver module for further downconversion and digital processing. The system ensures long-distance, low-loss signal transmission while preserving the integrity and polarization of the incoming astronomical signals.}
    \label{fig:RFsystem}
\end{figure}

The data acquisition terminal collects signals from 8 analog input channels, with each channel having a bandwidth of 128 MHz, an 8-bit resolution, and a recording data rate of about 16 Gbps. In this experiment, channels 0, 2, 4, and 6 receive signals from the same polarization direction of the four antennas, while channels 1, 3, 5, and 7 receive signals from the other polarization direction. The schematic of the RF analog system are shown in Fig.~\ref{fig:RFsystem}.

\clearpage

\clearpage
\subsubsection{CASSIOPEIA A OBSERVATION AND DATA PROCESSING}

\begin{figure}[H]
    \centering
    \includegraphics[width=0.9\linewidth]{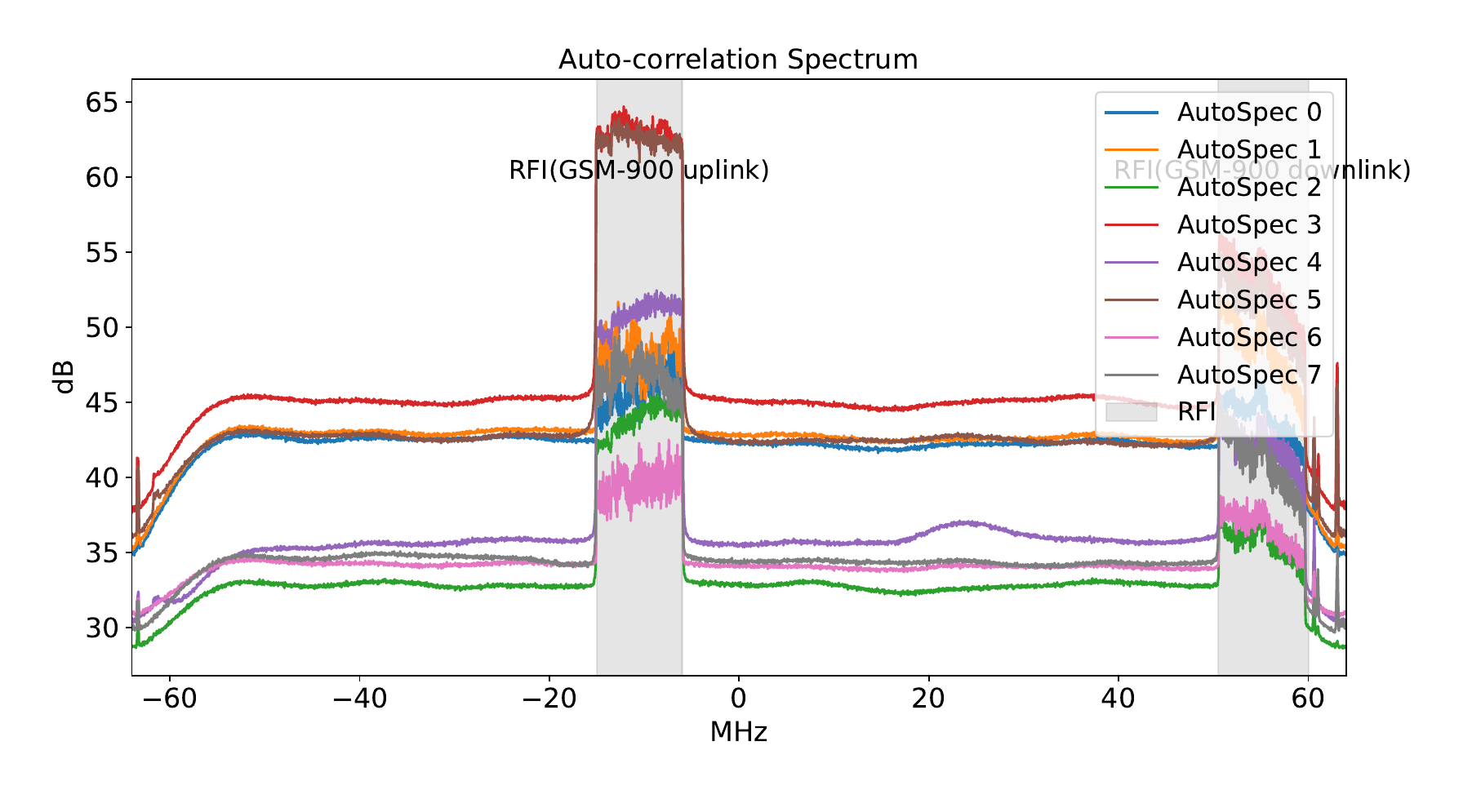}
    \caption{spectrum of Cassiopeia A in 8 channels(the center frequency is 896MHz)}
    \label{fig:autospectrum}
\end{figure}

\begin{figure}[H]
    \centering
    \includegraphics[width=0.9\linewidth]{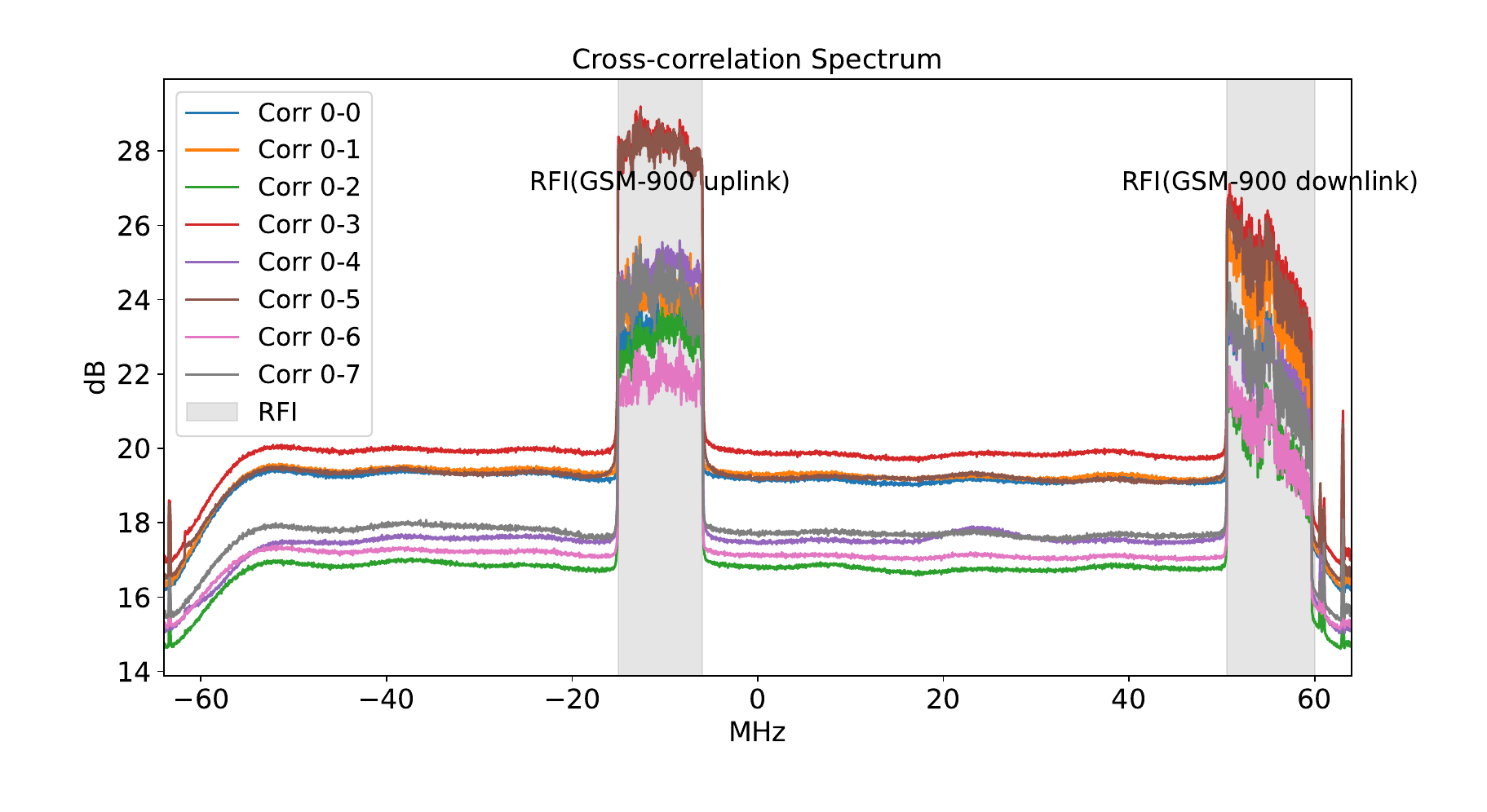}
    \caption{cross-correlation of Cassiopeia A among the eight signals(the center frequency is 896MHz)}
    \label{fig:cross_correlation spectrum}
\end{figure}

During the observation, data from the acquisition terminal was captured in real-time for online processing. Fig.~\ref{fig:autospectrum} shows the spectrum of the eight signals. Fig.~\ref{fig:cross_correlation spectrum} illustrates the cross-correlation among the eight signals. The horizontal axis represents the baseband frequency, with the center frequency of 0 MHz corresponding to a radio frequency of 896 MHz.

\subsubsection{waterfall plot}
\begin{figure}[H]
    \centering
    \includegraphics[width=1\linewidth]{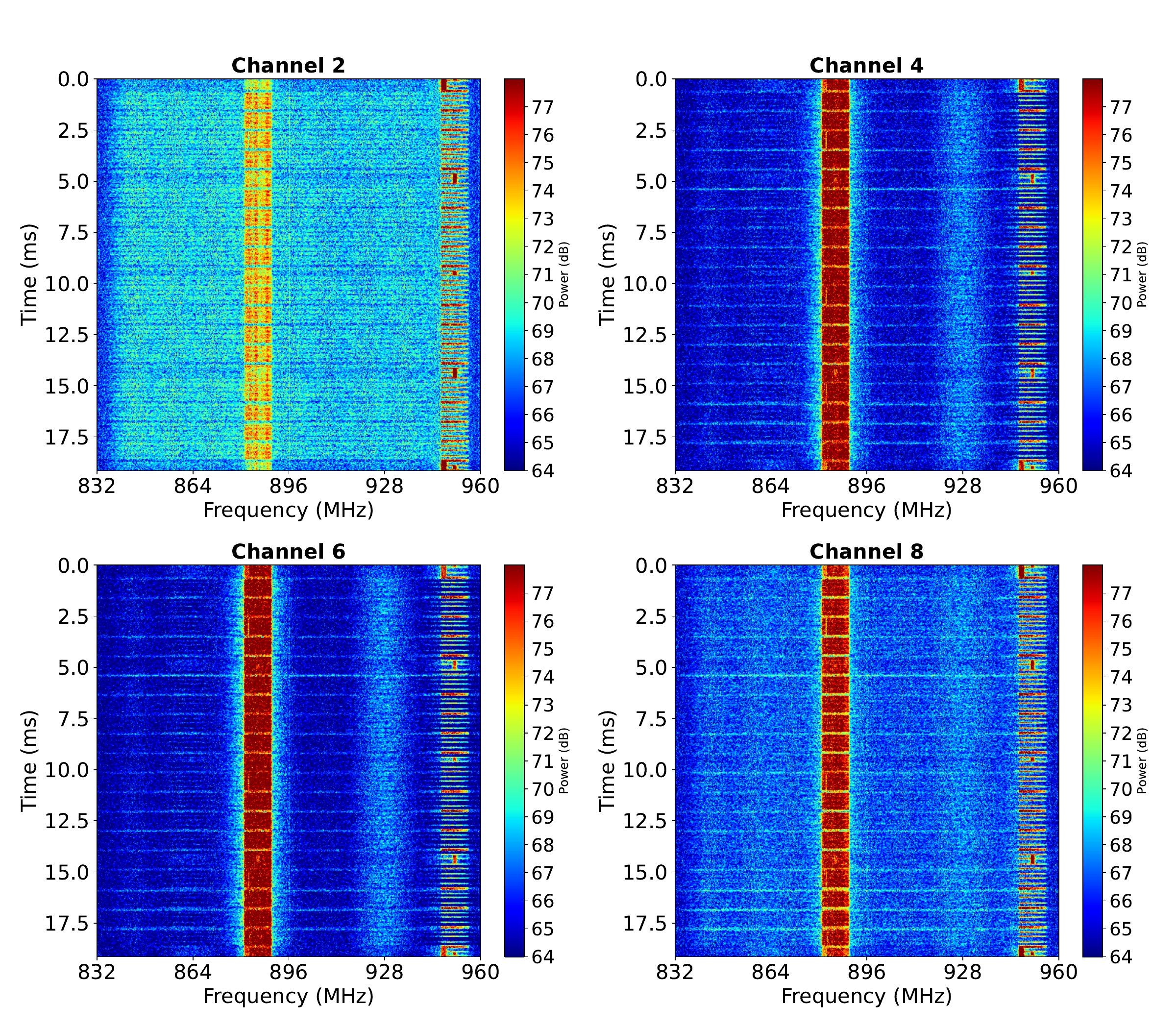}
    \caption{Waterfall plots of Channels 2, 4, 6, and 8, showing the time-frequency power distribution of the received signals.
The horizontal axis represents the frequency range from 832 to 960 MHz, while the vertical axis indicates time evolution over 18 milliseconds. The color scale denotes signal power in dB. Strong radio frequency interference (RFI) is clearly visible near 896 MHz, as well as periodic spurious signals beyond 930 MHz. These plots demonstrate the spectral and temporal behavior of the system, and highlight the presence of RFI features to be considered in later signal processing and delay estimation.}
    \label{fig:waterfall}
\end{figure}

Fig.~\ref{fig:waterfall} shows the waterfall plots of Channels 2, 4, 6, and 8 during the observation of Cassiopeia A. These time-frequency diagrams reveal the power distribution of the received signals, highlighting both the astronomical signal and the presence of radio frequency interference (RFI).

\section{time delay compensation}
\label{sec:delay_comp}
\subsection{interference fringes}
During the observation of Cassiopeia A, we observed clear interference fringes. The phase difference between antennas in a baseline is given by the following equation\citep{interference}:

\begin{equation}
    \Delta\phi = \frac{2\pi \mathbf{b} \cdot \hat{s}}{\lambda}
\end{equation}

where $\Delta\phi$ is the phase difference, $\mathbf{b}$ is the baseline vector between the two antennas, $\hat{s}$ is the unit vector in the direction of the source, and $\lambda$ is the observing wavelength. The phase difference is constrained within the range $[-\pi, \pi]$.

\begin{figure}[H]
    \centering
    \includegraphics[width=0.9\linewidth]{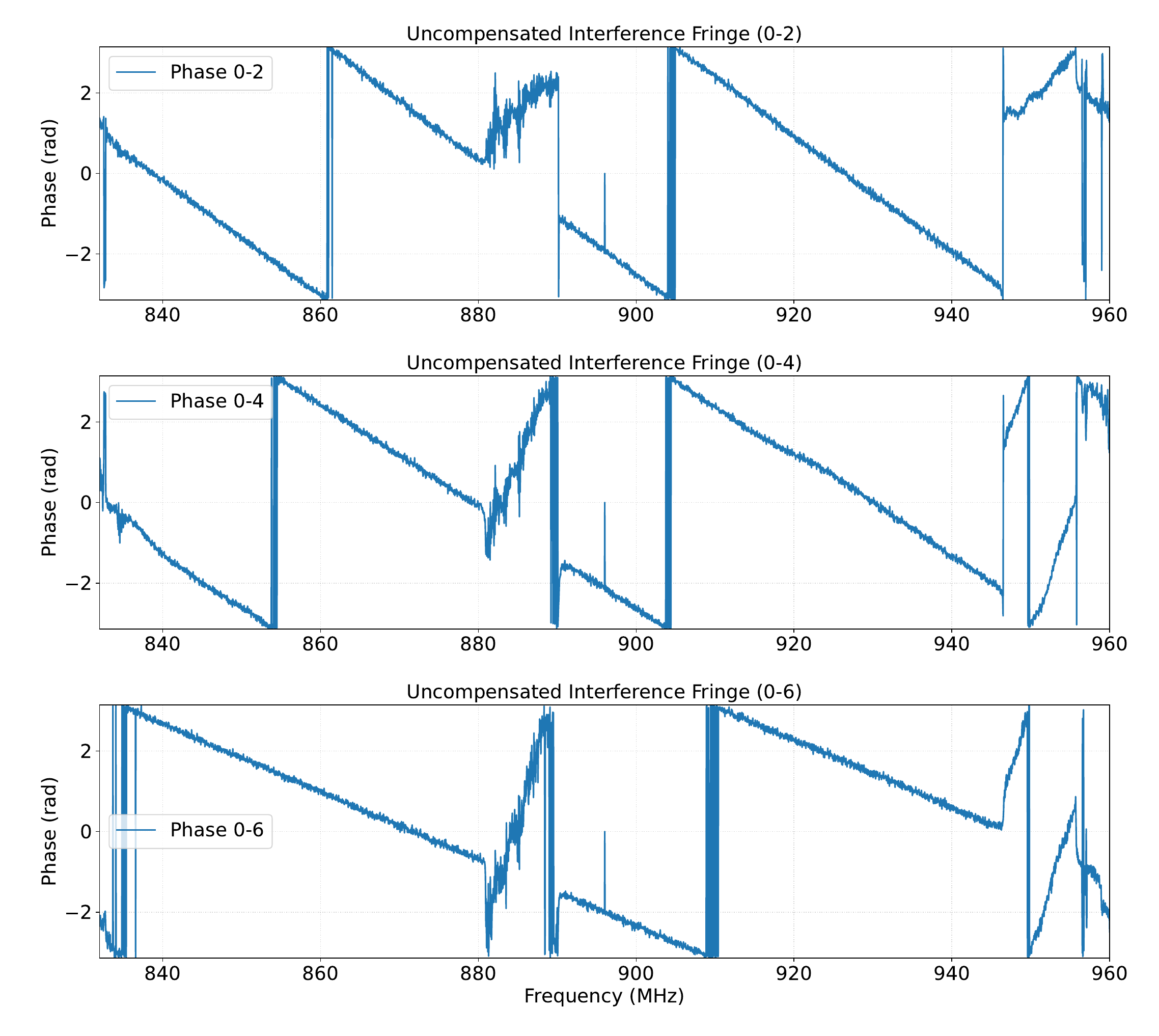}
    \caption{interference fringes of Cassiopeia A from Tianlai Array (the center frequency is 896MHz)}
    \label{fig:interference fringes}
\end{figure}

Based on this relationship, the resulting interference fringe pattern is shown in Fig.~\ref{fig:interference fringes}.

\subsection{Time Delay Estimation Methods}

\subsubsection{Geometric time delay estimation}
Based on the geometric configuration of our antenna array (baseline lengths of 12.04\,m, 22.63\,m, and 30.48\,m with corresponding angles of 55.89$^\circ$, 74.55$^\circ$, and 82.52$^\circ$), we calculated the theoretical relative delays between each antenna and the reference antenna \citep{vlbi}(Channel~0):

\begin{equation}
\tau_{ij} = \frac{b_{ij}\cos\theta_{ij}}{c}
\end{equation}

where $b_{ij}$ is the baseline length, $\theta_{ij}$ is the angle between the incident direction and the baseline, and $c$ is the speed of light. 

\subsubsection{Actual Time Delay Estimation via Gaussian Process Regression}

Based on the signal processing pipeline applied to our antenna array data, we estimated the actual relative delays between each antenna and the reference antenna (Channel~0) using cross-correlation and Gaussian Process Regression (GPR)\citep{GP,RGPR}. The estimation process involved the following steps:

\begin{equation}
\tau_{ij} = -\frac{1}{2\pi}\frac{d\phi_{ij}}{df}
\end{equation}

where $\phi_{ij}$ is the phase difference between channels $i$ and $j$, and $f$ is the frequency. The complete estimation procedure was:

\begin{enumerate}
\item Compute cross-correlation spectrum between reference and target channels:
\begin{equation}
S_{ij}(f) = \langle X_i(f)X_j^*(f) \rangle
\end{equation}

\item Extract phase difference:
\begin{equation}
\phi_{ij}(f) = \arg(S_{ij}(f))
\end{equation}

\item Apply robust phase unwrapping algorithm to remove $2\pi$ jumps

\item Select high-quality frequency points using:
\begin{itemize}
\item Amplitude threshold: $|S_{ij}(f)| > 70^{\text{th}}$ percentile
\item Frequency mask: exclude RFI-affected bands (878-897 MHz and 944-960 MHz)
\end{itemize}

\item Fit phase-frequency relationship using Gaussian Process Regression with kernel:
\begin{equation}
k(f,f') = C \cdot \text{RBF}(l=\SI{10}{MHz}) + \text{WhiteKernel}(\sigma^2=0.1)
\end{equation}

\item Compute delay from the mean slope of the fitted phase curve
\end{enumerate}

\subsection{Time delay Compensating Results}
\subsubsection{Geometric time delay estimation results}
The geometric time delay compensating results are shown in Tab.~\ref{tab:delay_params},
The computed delays are:
\begin{itemize}
\item Channel 0--2: $2.251 \times 10^{-8}$\,s
\item Channel 0--4: $2.010 \times 10^{-8}$\,s
\item Channel 0--6: $1.323 \times 10^{-8}$\,s
\end{itemize}

At our sampling rate of 128\,MHz, these delays correspond to:
\begin{itemize}
\item Channel 0--2: 2.881 sampling intervals
\item Channel 0--4: 2.572 sampling intervals
\item Channel 0--6: 1.693 sampling intervals
\end{itemize}

\begin{table}[H]
\centering
\caption{Theoretical time delay compensation parameters}
\label{tab:delay_params}
\begin{tabular}{ccccc}
\hline
Channel Pair & Theoretical Delay (s) & Sampling Intervals & Integer Comp. & Fractional Comp. \\
\hline
0--2 & $2.251 \times 10^{-8}$ & 2.881 & 2 & 0.881 \\
0--4 & $2.010 \times 10^{-8}$ & 2.572 & 2 & 0.572 \\
0--6 & $1.323 \times 10^{-8}$ & 1.693 & 1 & 0.693 \\
\hline
\end{tabular}
\end{table}

\subsubsection{Actual time delay estimation results}
\begin{figure}[h]
    \centering
    \includegraphics[width=1.0\linewidth]{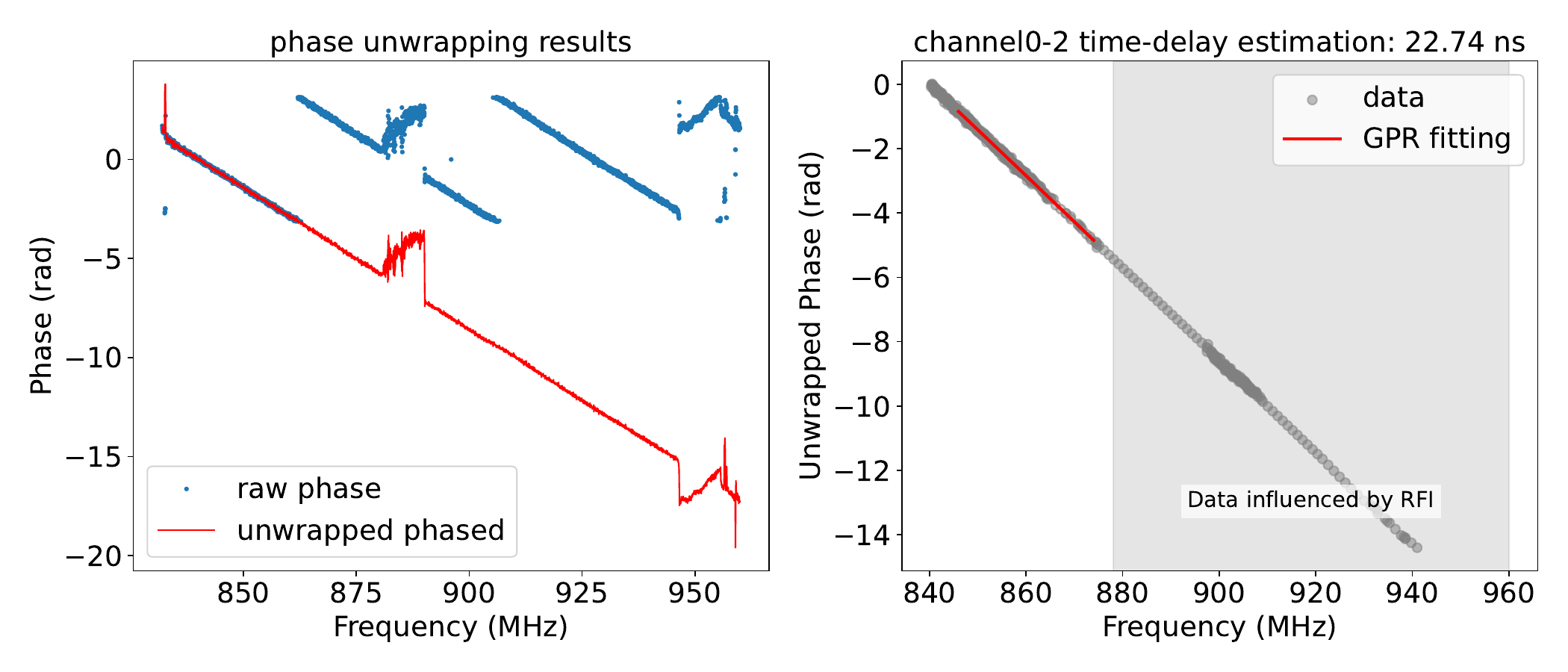}
    \caption{An example of the Gaussian Process Regression (GPR) phase slope fitting process for Channel 0–2. The left panel shows a comparison between the original interference fringes and the unwrapped phase after $2\pi$ correction. The right panel displays the phase slope fitted by GPR, with the RFI-contaminated frequency regions excluded. The resulting phase slope is used to estimate the actual time delay between the channels.
}
    \label{fig:GP}
\end{figure}

The measured actual delays and their corresponding sampling intervals at our \SI{128}{MHz} sampling rate are:

\begin{itemize}
\item Channel 0--2: $\SI{2.269e-8}{s}$ (\SI{2.91}{sampling\ intervals})
\item Channel 0--4: $\SI{1.977e-8}{s}$ (\SI{2.53}{sampling\ intervals})
\item Channel 0--6: $\SI{1.317e-8}{s}$ (\SI{1.69}{sampling\ intervals})
\end{itemize}
The Gaussian Process fitting procedure is visualized in Fig.~\ref{fig:GP}.

\subsubsection{Comparison between theoretical values and actual values}

As is shown in Tab.~\ref{tab:comparison}, the actual delays show excellent agreement with the theoretical predictions calculated from the array geometry:

\begin{table}[H]
\centering
\begin{tabular}{lcc}
\hline
Channel Pair & Theoretical Delay & Actual Delay \\
\hline
0--2 & \SI{2.251e-8}{s} & \SI{2.269e-8}{s} \\
0--4 & \SI{2.010e-8}{s} & \SI{1.977e-8}{s} \\
0--6 & \SI{1.323e-8}{s} & \SI{1.317e-8}{s} \\
\hline
\end{tabular}
\caption{Comparison between theoretical and actual delays}
\label{tab:comparison}
\end{table}

The small discrepancies (typically within \SI{5}{\%}) between the theoretical geometric delays and the actual delays estimated from signal processing can be attributed to the following factors:

\begin{itemize}
\item \textbf{Antenna Positioning Errors:} Small coordinate errors in antenna placement can cause baseline inaccuracies that propagate into delay estimates.\citep{interference}

\item \textbf{Atmospheric and Ionospheric Effects:} Residual tropospheric and ionospheric delays may cause errors, especially at low elevation angles.\citep{GNSS}

\item \textbf{Residual Radio Frequency Interference (RFI):} Undetected or low-level RFI can distort phase slope fitting and bias delay estimation.\citep{2013lofar}

\end{itemize}

\subsection{Implementation of Delay Compensation}
\label{subsec:implementation}

We implemented a two-stage compensation scheme\citep{vlbi}:

\begin{enumerate}
\item \textbf{Integer delay compensation:}
\begin{equation}
\Delta n_{\mathrm{int}} = \lfloor \tau_{ij}f_s \rfloor
\end{equation}
implemented through circular shift:
\begin{equation}
x'[n] = x[n-\Delta n_{\mathrm{int}}]
\end{equation}

\item \textbf{Fractional delay compensation:}
For the remaining fractional parts ($\Delta n_{\mathrm{frac}} = 0.91$, 0.53, and 0.69 for Channel pairs 0--2, 0--4, and 0--6 respectively), we applied time-domain interpolation:
\begin{equation}
x''[n] = \mathrm{interp1d}(x'[n], \Delta n_{\mathrm{frac}})
\end{equation}
using linear interpolation to maintain phase continuity.
\end{enumerate}

Based on the phase slope fitting using Gaussian Process Regression (GPR), the actual time delays between each antenna and the reference channel were estimated from the unwrapped cross-correlation phase spectra. These delays incorporate the effects of instrumental chain, array geometry, and residual RFI contamination.

The purpose of this delay compensation is to correct signal arrival time differences based on measured phase behavior rather than purely geometric assumptions. This improves fringe coherence and visibility, ensuring more accurate interferometric measurements of the source structure.

\begin{figure}[H]
    \centering
    \includegraphics[width=0.7\linewidth]{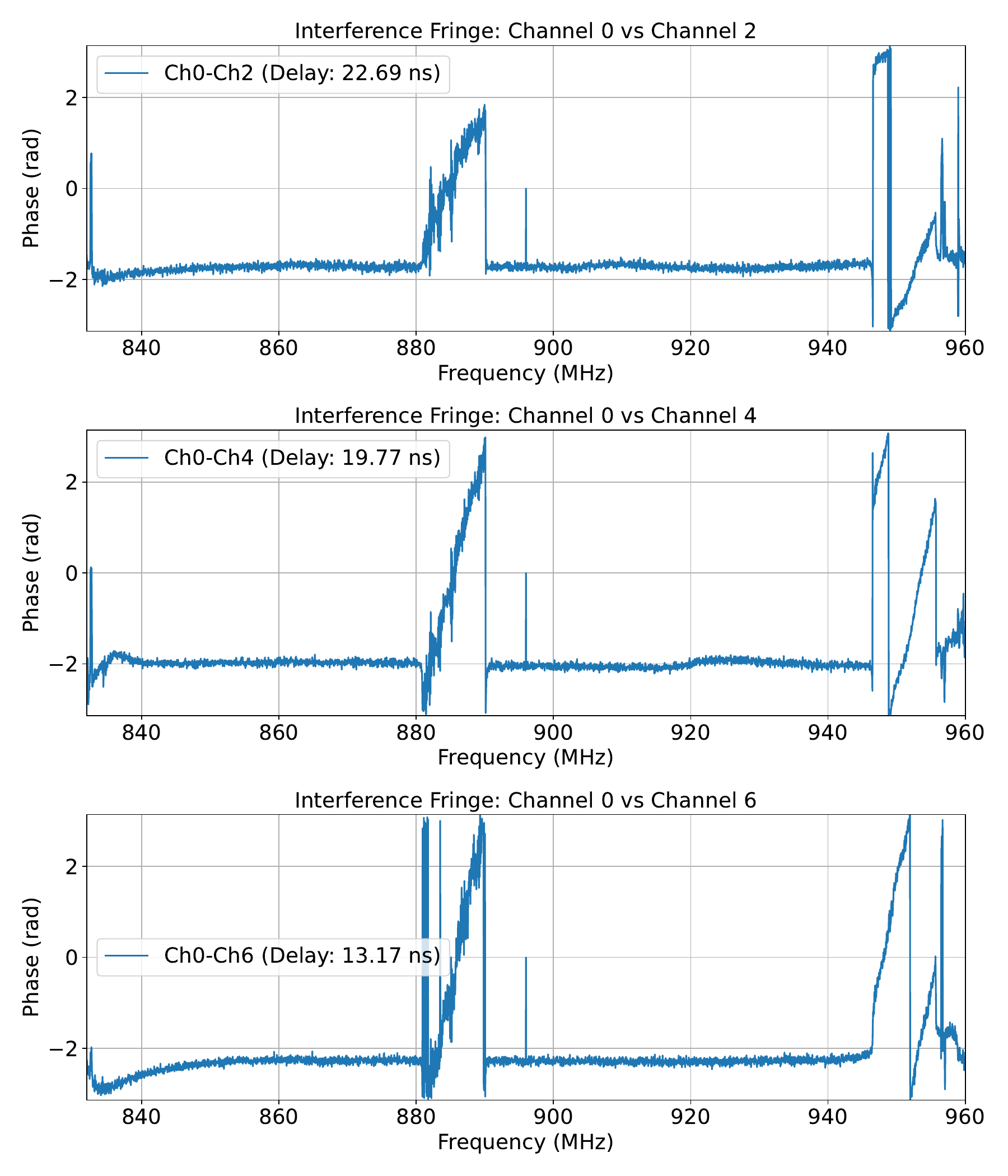}
    \caption{The figure shows the interference fringes of Cassiopeia A after time delay compensation. After applying the compensation, the phase slope becomes zero, indicating that the time delay between the two antennas has been corrected. The overall phase offset is attributed to the intrinsic phase difference between the antennas.
}
    \label{fig:interference fringes comp}
\end{figure}

Using the GPR-estimated delays, we applied corresponding integer and fractional delay compensation to the signal streams. The resulting interference fringes, shown in Fig.~\ref{fig:interference fringes comp}, demonstrate significant phase flattening and enhanced coherence between antennas.

\begin{figure}[H]
    \centering
    \includegraphics[width=0.7\linewidth]{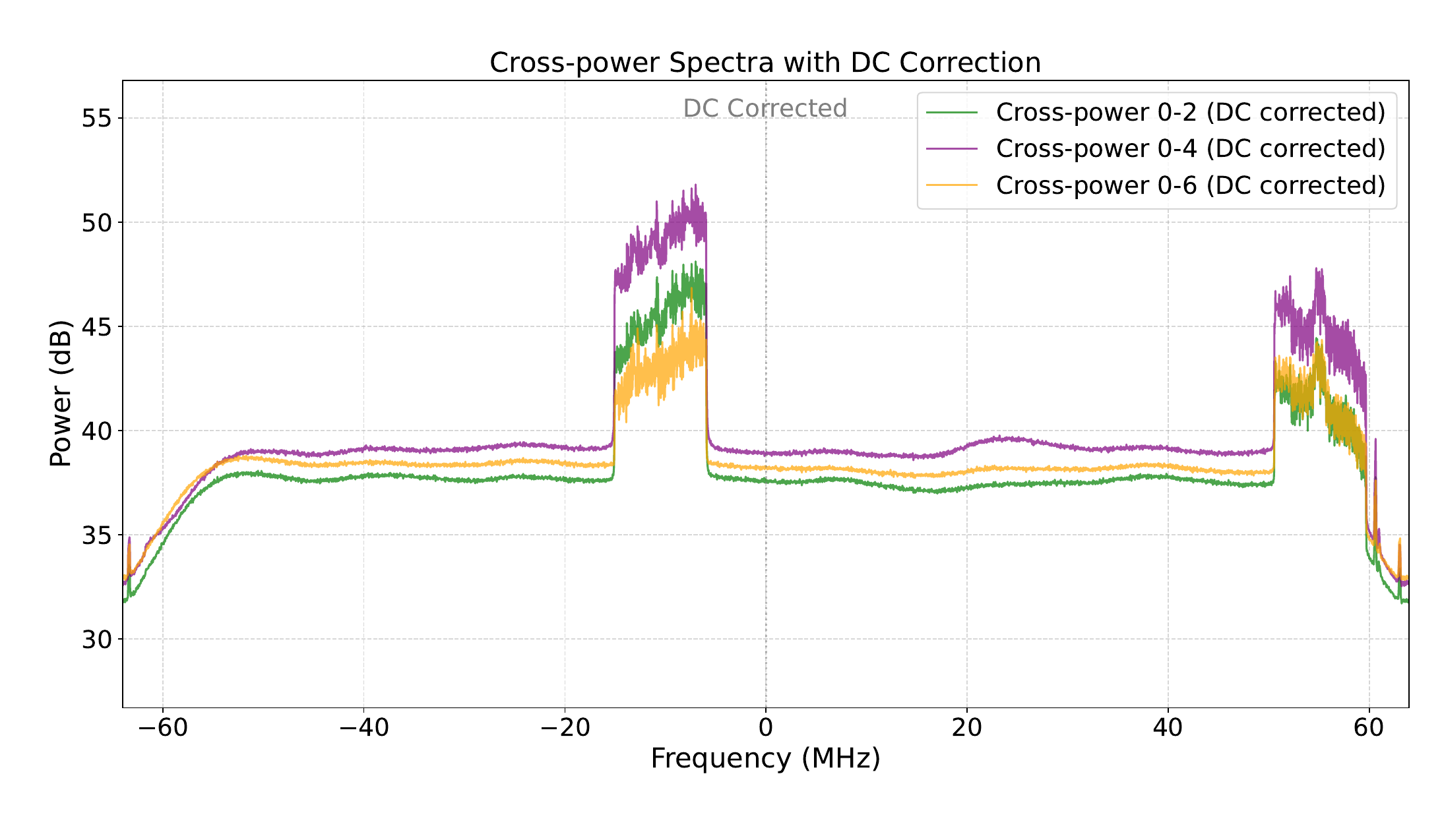}
    \caption{The figure presents the cross-correlation spectrum after applying time delay compensation. The overall signal amplitude is noticeably enhanced compared to the uncompensated case, indicating improved coherence between the channels.
}
    \label{fig:cross-correlation-compensation}
\end{figure}

The cross-correlation power spectra of Cassiopeia A after applying time delay compensation are shown in Fig.~\ref{fig:cross-correlation-compensation}.

\section{Conclusion and discussion}
\label{sec:conclusion}
In this work, we designed, implemented, and validated a high-speed digital receiver system intended for use in the BINGO/ABDUS interferometric platform. The receiver supports 8-channel analog input at 12-bit resolution and up to 4~GHz sampling, enabling real-time channelization and data acquisition for radio astronomy applications. Field observations using the Xi’an and Tianlai dish arrays were conducted to evaluate the system's performance.

Through correlation analysis of signals from both artificial and celestial sources—including a satellite beacon, the Sun, and Cassiopeia A—we successfully detected interference fringes and conducted detailed time delay calibration. We applied a two-stage delay compensation scheme combining integer-sample shifting and fractional interpolation. The time delays were estimated both theoretically (based on antenna baseline geometry) and empirically via Gaussian Process Regression (GPR) of the cross-correlation phase slopes.

The comparison between geometric and fitted delay values showed excellent agreement within 5\%, validating both the analog and digital subsystems. After delay compensation, improved phase alignment and enhanced cross-correlation amplitudes confirmed the effectiveness of the system for interferometric imaging and calibration.

These results confirm that the digital backend is suitable for phased array and outrigger deployment in the BINGO/ABDUS system. The receiver system's architecture also demonstrates scalability for future use in large-N interferometric arrays. Future work will focus on full-array beamforming, automated delay tracking, polarization calibration, and on-sky interferometric imaging with a larger number of antennas.

\section{acknowledgment}
YW acknowledges the support from the National SKA Program of China(2020SKA0110103).
JJZ acknowledges the support from NSFC grant No.12473003.
We acknowledge the support from Yifeng Li, Jintao Luo and Fengquan Wu for conducting the experiments.
\bibliographystyle{raa}  
\bibliography{bibtex}

\end{document}